\documentclass[prb,twocolumn,showpacs,preprintnumbers,amsmath,amssymb]{revtex4}
\usepackage{graphicx}
\usepackage{dcolumn}
\usepackage{bm}

\def\bk{ {\bf k} }
\def\bq{ {\bf q} }
\def\bd{ {\bf d} }
\def\be{ {\bf e_z} }

\def\re{ \,{\rm Re}\, }

\begin{document}

\title{Determining the superconducting gap structure in Sr$_2$RuO$_4$
from sound attenuation studies below $T_c$}

\author{Pedro Contreras}
\author{Michael Walker}
\affiliation{Department of Physics, University of Toronto,
Toronto, ON M5S 1A7, Canada}
\author{Kirill Samokhin}
\affiliation{Department of Physics, Brock University, St.
Catharines, ON L2S 3A1, Canada}

\date{\today}

\begin{abstract}
This work presents a quantitative theoretical study of the sound
attenuation in the unconventional multiband superconductor
Sr$_2$RuO$_4$ below the superconducting transition temperature
T$_c$.  Sound attenuation in this material is shown to have the
remarkable property of being able to identify different nodal
structures on different bands. The nodal structures on the
$\gamma$ band on the one hand, and on the $\alpha$ and $\beta$
bands on the other, are both found to be characterized by the
existence of point nodes, but are significantly different in their
quantitative aspects.
\end{abstract}

\maketitle

\section{Introduction}

Sr$_2$RuO$_4$ is an unconventional superconductor which has a
Fermi surface composed of three sheets, called the $\alpha,
\beta,$ and $\gamma$ sheets (i.e. it is an example of a multiband
superconductor).  There is considerable agreement as to the
symmetry of the superconducting state, but there is as yet no
consensus as to the nodal structure of the superconducting gap on
the different sheets of the Fermi surface.  The purpose of this
article is first of all to develop a model of a possible nodal
structure, and secondly to present a detailed quantitative
analysis of an ultrasonic attenuation
experiment\protect\cite{lup01} which confirms the essential
features of the model. It is a remarkable feature of ultrasonic
attenuation in Sr$_2$RuO$_4$ that it is able to distinguish
different nodal structures on different sheets of the Fermi
surface. It does this because different sound wave modes are
sensitive to quasiparticles on different parts of the Fermi
surface.

Sr$_2$RuO$_4$ has a layered square-lattice structure similar to
that of many high-temperature copper-oxide superconductors
\protect\cite{mae94}. The critical temperature varies strongly
with impurity concentration, and  can be as high as $T_c \approx$
1.5 K. Furthermore, the normal state displays Fermi liquid
behavior. (see Refs.~\protect\onlinecite{mae01} and
\protect\onlinecite{mae02} for reviews of the properties of
Sr$_2$RuO$_4$.)  There is currently a reasonable consensus that in
Sr$_2$RuO$_4$ the Cooper pairs form in a spin-triplet state which
breaks time reversal symmetry \protect\cite{ric95,luk98,walk3},
and that the broken time-reversal symmetry results from a
two-component order parameter of E$_{u}$ symmetry.

There is, however, no consensus concerning the detailed structure
of the nodes in the superconducting gap.  The current situation is
reviewed in Ref.~\protect\onlinecite{mae02} where two promising
scenarios are discussed. Both scenarios involve line nodes, which
would give a low-temperature specific heat varying approximately
linearly with temperature, in qualitative agreement with the
observations of Ref.~\protect\onlinecite{nis99}. In one scenario,
horizontal line nodes in the superconducting gap are proposed
\protect\cite{yasu00,zhi01,wys03} (note however that such
horizontal line nodes are not stable and tend to become point
nodes \protect\cite{ber03}). Qualitative arguments indicate that
horizontal line nodes can be consistent with ultrasonic
attenuation measurements.\protect\cite{lup01,wak01} Whether or not
the angular dependence of the thermal conductivity measured in a
magnetic field also provides evidence for horizontal line nodes is
discussed in Ref.~\protect\onlinecite{mae02}.  Another potential
scenario \protect\cite{miy99} is the presence of a very small gap
along the [100] directions where the Fermi surface is close to the
zone boundary.  While such depressions of the gap in the [100]
directions may exist, it is clear that these can not be
responsible for the observed ultrasonic attenuation. General
symmetry arguments (independent of model details) show that the
interaction of transverse [100] phonons with quasiparticles having
wavevectors in a [100] direction is zero.\protect\cite{wak01} Thus
nodal structures not along the [100] direction must be present to
explain ultrasonic attenuation measurement.

In Section II we give a detailed analysis of the nodal structure
expected for a broken time-reversal symmetry state of triplet
E$_{u}$ symmetry; this analysis differs considerably from those
given previously.  Here, an important result is anticipated using
a general argument. The starting point is the assumption of gap
characterized by a d-vector of the form
\begin{equation}
    \label{new_repr_gap}
        \bd^{i} ( \bk ) = \be \Big( d_x^i (\bk) + i ~d_y^i (\bk)
        \Big),
\end{equation}
where the superscript $i$ refers to the $\alpha, \beta,$ or
$\gamma$ part of the Fermi surface (see Fig.~\ref{nodes} below)
and where $d_x^i$ and $d_y^i$ are real and are the two components
of the order parameter on Fermi surface sheet $i$. The energies of
the Bogoliubov quasiparticle excitations corresponding to this
d-vector are
\begin{equation}
        \label{sup_spectrum}
            E^i(\bk) = \sqrt{(\epsilon^i_{\bf k})^2
                + [\Delta^i(T)]^2
                    \{[d_x^i ({\bf k})]^2 + [d_y^i ({\bf k})]^2\}}.
\end{equation}
The nodal points, i.e. the points where the Bogoliubov
quasiparticle energy is zero, determine the low-temperature
thermodynamic properties.\protect\cite{book} These points are
determined by the equations
\begin{equation}
    \label{nodes_general}
            \epsilon^i_{\bf k} = 0,\ \ \ d_x^i ({\bf k}) = 0,
                \ \ \  d_y^i ({\bf k}) = 0.
\end{equation}
For a given $i$, each of these three equations represents a
surface in the three-dimensional ${\bf k}$ space. In general, two
surfaces intersect, if they intersect at all, on a line. If a
third surface intersects this line it will intersect it at a
point. Thus, in general, the solution of the three
Eqs.~\ref{nodes_general} for a given $i$ will be a point. Hence,
we should expect point nodes, if any, for a broken time-reversal
symmetry state such as is being proposed for Sr$_2$RuO$_4$. Lines
nodes might occur for very special values of the material
parameters, but this would be a highly unusual occurrence and
should not be expected.

As noted above, the main point of this article is to develop a
detailed quantitative theory of the ultrasonic attenuation in
Sr$_2$RuO$_4$ and to extract the details of the nodal structure of
Sr$_2$RuO$_4$ from a comparison of the results of this calculation
with the experimental ultrasonic attenuation results of
Ref.~\protect\onlinecite{lup01}. Some details of the
electron-phonon interaction important for the calculation of
ultrasonic attenuation are given at the end of Section II. It is
of particular importance to recall that the
measured\protect\cite{lup01} ultrasonic attenuation in
Sr$_2$RuO$_4$ is exceptional in its extreme anisotropy.  This
extreme anisotropy has been shown in
Ref.~\protect\onlinecite{wak01} to be due to the layered
square-lattice structure of Sr$_2$RuO$_4$, and to occur only in
the interaction of phonons with electrons in the $\gamma$ band,
but not with electrons in the $\alpha$ and $\beta$ bands. It is
this fact that allows the attenuation of the three modes,
identified in Ref.~\protect\onlinecite{lup01} as the L[100],
L[110], and T[110] modes, as being due to the interaction with
$\gamma$-band quasiparticles. The approximate $T^2$
low-temperature variation of ultrasonic
attenuation\protect\cite{lup01} of these three modes suggests that
the attenuation is due to $\gamma$-band point nodes, a result that
is confirmed in more detailed numerical calculations. (In general,
line nodes and point nodes give an ultrasonic attenuation that
varies as $T$ and $T^2$ respectively in the low temperature limit
-- see Section III).

The electron-phonon interaction responsible for the attenuation of
the L[100], L[110], and T[110] phonon modes turns out, however, to
be exactly zero for the T[100] phonon mode, and the attenuation
for this phonon mode has a different temperature dependence
($T^{1.4}$) from that of the other three modes. This suggests that
the attenuation of the T[100] mode is due to interactions with
quasiparticle in a different band (the $\alpha$ or $\beta$ bands,
or both).  The model nodal structure that fits these is also a
structure with point nodes, but with a very low gap on a
horizontal line joining the point nodes.  Thus the $\alpha$-
and/or $\beta$-band nodal structure would appear to be
line-node-like at temperatures which are not too low, and would
dominate the overall thermodynamic behavior. The model proposed
here would therefore not be inconsistent with the approximately
linear in $T$ behavior of the specific heat observed in
Ref.~\protect\onlinecite{nis99}.

\begin{figure}
\includegraphics[width = 3.2 in, height= 3.2 in]{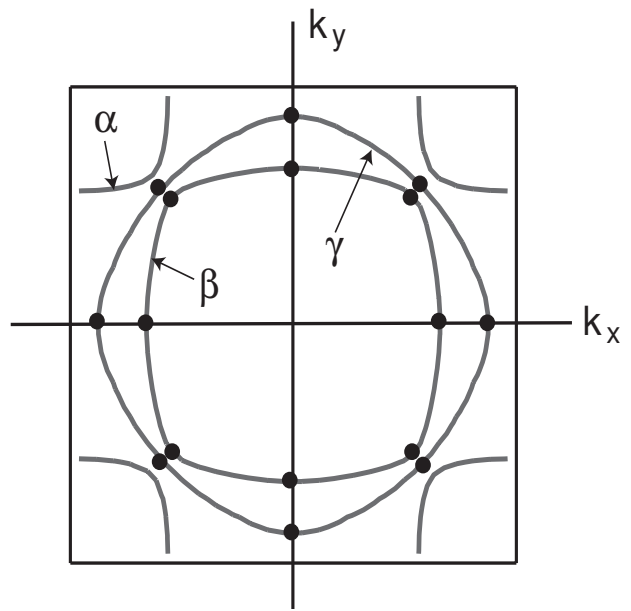}
\caption{\label{nodes} The solid circles show the positions of the
point nodes in the superconducting gap on the $\beta$ and $\gamma$
Fermi surface sheets in Sr$_2$RuO$_4$, as determined by
Eqs.~\ref{real_gap} and \ref{imag_gap}. Each solid circle
represents two nodes, at positions $\pm k_z$. Evidence is
presented in the article for the existence of point nodes in the
$\gamma$ band, and in the $\alpha$ or $\beta$ bands, or both.}
\end{figure}

Section III gives details of the formulae used to calculate the
ultrasonic attenuation as a function of temperature. Section IV
compares the results of our numerical calculations with the
experimental results and confirms the qualitative statements of
the preceding paragraph. The two previous studies most closely
related to ours are those of Refs.~\protect\onlinecite{gra01}, and
\protect\onlinecite{wu02}. Both are carried out using isotropic
models, as are all the previous studies of ultrasound attenuation
in superconductors that we are aware of. However as noted in
Ref.~\protect\onlinecite{wak01} isotropic models can give a
misleading idea of the position of the gap nodes in unconventional
superconductors. Thus, the extension of the treatment of
ultrasonic attenuation to the anisotropic case, as is done below,
is essential for a correct identification of the nodal structure.

It has been shown some time ago \protect\cite{var86} that the
transport and thermal properties of heavy-fermion superconductors
can be explained in terms of an effective electron scattering rate
which except for the lowest temperatures, is approximately
temperature independent and equal in magnitude to that of the
normal state. Such a lifetime arises in a self-consistent
treatment of impurity scattering near the unitary limit. Thus, in
our treatment which is not self-consistent, we take the
superconducting quasiparticle lifetime to be a constant and equal
to that in the normal state (both being assumed to be due to
impurity scattering). This approach has been found to be
successful in ultrasonic attenuation studies in UPt$_3$, (see
Ref.~\protect\onlinecite{gra02}).  A full self-consistent
treatment solving for the anisotropic multiband gap function and
energy-dependent scattering rate is beyond the scope of this
article.

\section{Details of the superconducting gap and the electron-phonon
interaction in $\mathrm{\bf {Sr_2RuO_4}}$}

The energy of a normal-state $i$-band electron  ($i = \alpha,
\beta, \gamma$) is periodic in reciprocal space, i. e. $\epsilon^i
(\bk)$ = $\epsilon^i (\bk + \bf{G})$ where $\bf{G}$ is any
reciprocal lattice vector. This means that $\epsilon^i (\bk)$ can
be written as a lattice Fourier series in the form
\begin{eqnarray}
        \label{energy_lattice}
            \epsilon^i (\bk) & = & \sum_{n}
                \epsilon^i_n \; e^{ i \; \bk \cdot \bf{R_n}},
\end{eqnarray}
where the sum is over all of the vectors $\bf{R_n}$ of the Bravais
lattice. The particular form of this expression used in this
article for the $\gamma$-band energy is the so called $t-t'$
approximation, (eg. see Ref.~\protect\onlinecite{maz97}). In this
approximation, electrons are assumed to hop between the Ru ions of
a single layer of the Sr$_2$RuO$_4$ structure. Hopping between the
layers is not allowed. The permitted intralayer hoppings are
either nearest neighbor (matrix element $t$) or next-nearest
neighbor (matrix element $t'$). This gives
\begin{eqnarray}
    \epsilon^{\gamma}_{\bf k} & = & E_0 + 2t(\cos(k_xa) + \cos(k_ya)) + \nonumber \\
                          & & 4t^\prime \cos(k_xa)\cos(k_ya).
\end{eqnarray}
The values for the $\gamma$-band tight binding parameters used in
this paper are $(E_0 - E_F, t, t^\prime)$ = $(-0.4, -0.4, -0.12)$

In a similar manner, the $d$-vector describing the superconducting
gap in band $i$ can be expanded as
\begin{eqnarray}
    \label{gap_lattice}
        {\bf d}^i(\bk) & =  &\sum_{n} {\bf d}^i_n \; e^{ i \bk \cdot {\bf R}_n}.
\end{eqnarray}
Here we adopt the view of Refs.~\protect\onlinecite{mae02,mae01}
that the order parameter transforms as the E$_{u}$ irreducible
representation of the point group D$_{4h}$ of Sr$_2$RuO$_4$, and
that the state of E$_{u}$ symmetry that is realized is one of
broken time-reversal symmetry. The appropriate form of the
$d$-vector is then given by Eq.~\ref{new_repr_gap}. The explicit
expressions that we use for $d_x^i$ and $d_y^i$ are
\begin{equation}
        \label{real_gap}
            d_x^i (\bk) =\delta^i\sin (k_x a) + \sin(\frac{k_x a}{2})
                \cos(\frac{k_y a}{2})\cos(\frac{k_z c}{2}),
\end{equation}
and
\begin{equation}
        \label{imag_gap}
            d_y^i (\bk)=  \delta^i\sin (k_y a) + \cos(\frac{k_x a}{2})
                \sin(\frac{k_y a}{2})\cos(\frac{k_z c}{2}).
\end{equation}
There are two different sets of basis vectors for the E$_{u}$
representation occurring in Eqs.~\ref{real_gap} and
~\ref{imag_gap}, one set contains the factor $\delta^i$, and the
second set contains the factor $\cos(k_z c/2)$. Related
expressions for the $d$-vector have been given previously in
Refs.~\protect\onlinecite{yasu00,zhi01,wys03}, but always in
combinations that give horizontal line nodes or no nodes. For
example all the authors consider the case $\delta^i$=0, which has
horizontal line nodes at k$_z$ = $\pm$ $\pi/c$. However, any
non-zero $\delta^i$ will remove the line nodes and produce instead
point nodes. In order to fit the experimental results on
ultrasound attenuation we will need quite a substantial value of
$\delta^\gamma$ for the $\gamma$ band and the point nodes in the
$\gamma$ band will play an important role. For the $\alpha$ and
$\beta$ bands the model of Eqs.~\ref{real_gap} and \ref{imag_gap}
holds also, but it will be shown that $\delta^\alpha$ or
$\delta^\beta$, or both, must be relatively small, but non-zero.

To analyze in detail the nodal structure for Sr$_2$RuO$_4$, note
that the Bogoliubov quasiparticle energy is given by
Eq.~\ref{sup_spectrum}. Clearly, for the superconducting energy
gap to be zero on a given sheet $i$ of the Fermi surface, the
three Eqs.~\ref{nodes_general} must be satisfied. For non-zero
$\delta^i$ in a certain range of values these equations have
solutions, but only for $\bk$ lying on the symmetry equivalent
$\lbrace100\rbrace$ and $\lbrace110\rbrace$ planes. Assuming a
solution exists, there are eight symmetry-related nodes in
$\lbrace100\rbrace$ planes (see Fig.~\ref{nodes})for a given $i$;
two of these nodal points are given by
\begin{equation}
        \label{nodes100}
            k_z c = \pm 2 \cos^{-1}[-2 \delta^i \cos(k_x a/2)]
\end{equation}
where $\bk$ = $(k_x,0,k_z)$ lies on the Fermi surface. There are
also eight symmetry-related nodes in $\lbrace110\rbrace$ planes
for a given $i$; two of these nodal points are given by
\begin{equation}
        \label{nodes110}
            k_z c = \pm 2 \cos^{-1}(-2 \delta^i)
\end{equation}
where $\bk$ = $(k_x,k_x,k_z)$ lies on the Fermi surface as well
(see Fig.~\ref{nodes} for details).

\begin{figure}
\includegraphics[width = 2.4 in, height= 2.6 in]{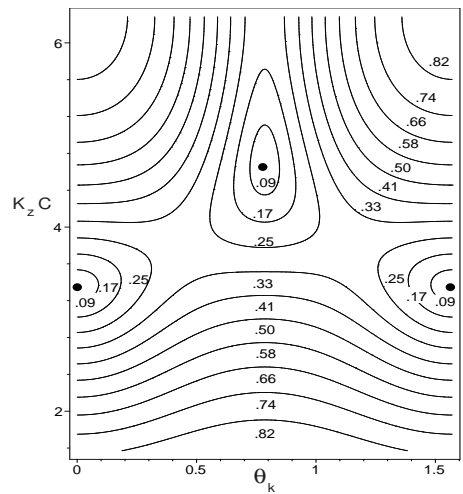}
\caption{\label{location1} Contour plot of the gap magnitude
$|d^\gamma_{\bf k}|$ on the Fermi surface as a function of the
longitudinal angle $\theta_{\bf k}$  relative to the [100] plane
(see text), and $k_z c$ ($k_z c$ varies from $\pi/2$ to $2 \,
\pi$). The solid circles indicate the nodal points where the gap
is zero.}
\end{figure}

For these solutions to exist, we must have $\vert 2 \delta^i
\cos(k_x a/2) \vert$ $\leq$ 1 in the first case, and $\vert
2\delta^i \vert$ $\leq$ 1 in the second. Clearly, the existence of
nodes in the second case implies their existence in the first
case, but the nodes of the first case may or may not be
accompanied by those of the second case. These point nodes are
``accidental'' in the sense that they are not required by
symmetry, but exist only if the material parameters have values in
a certain range. Also, these point nodes will degenerate into the
line nodes discussed by previous authors
\protect\cite{zhi01,yasu00,wys03} if $\delta^i$ is exactly zero,
but this should not in general be expected to be the case.

\begin{figure}
\includegraphics[width = 2.4 in, height= 2.6 in]{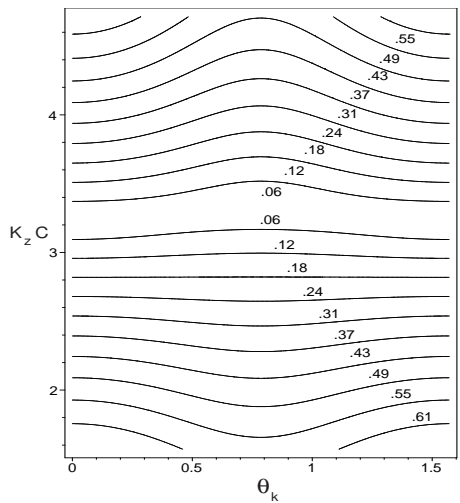}
\caption{\label{location2} Contour plot of the gap magnitude
$|d^\beta_{\bf k}|$ on the Fermi surface as a function of the
longitudinal angle $\theta_{\bf k}$  relative to the [100] plane
(see text), and $k_z c$ ($k_z c$ varies from $\pi/2$ to $3/2 \,
\pi$).}
\end{figure}

\begin{figure}
\includegraphics[width = 2.4 in, height= 2.6 in]{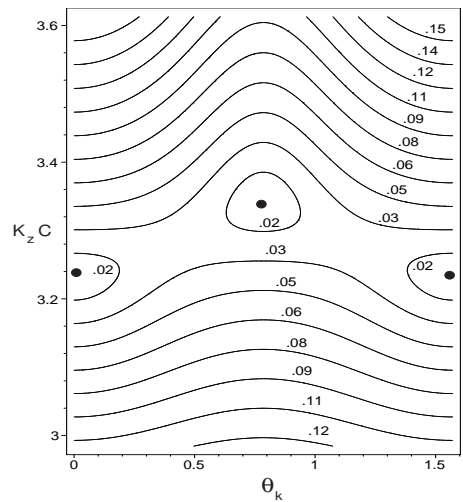}
\caption{\label{location3} Contour plot of the gap magnitude
$|d^\beta_{\bf k}|$ on the Fermi surface as a function of the
longitudinal angle $\theta_{\bf k}$  relative to the [100] plane
(see text), and $k_z c$ ($k_z c$ varies from $0.95 \pi$ to $1.15
\, \pi$). The solid circles indicate the nodal points where the
gap is zero.}
\end{figure}

Contour plots of
\begin{equation}
    |d^i_{\bf k}| = ([d^i_x({\bf k})]^2 + [d^i_y{\bf k})]^2)^{1/2}
\end{equation}
on the Fermi surface for both the $\gamma$ and $\beta$ bands are
shown in Figs.~\ref{location1}, \ref{location2}, and
\ref{location3}. This quantity is proportional to the magnitude of
the gap at wave vector ${\bf k}$. To obtain a relatively simple
graphical view of the nodal structure, the Fermi surfaces are
parameterized using $\bk$ = $(\pi R \cos \theta_{\bk}, \pi R \sin
\theta_{\bk},k_z)$ with $R = 0.9$ for the $\gamma$ band, and
$R=0.7$ for the $\beta$ band [i.e. for this calculation the Fermi
surfaces are approximated by right circular cylinders (c.f.
Ref.~\protect\onlinecite{miy99})]. The values $\delta^\gamma =
0.35$ and $\delta^\beta = 0.05$, determined in Section IV by
fitting the ultrasonic attenuation measurements to experiment, are
used.

Notice (Fig.~\ref{location1}) that the $\gamma$ band has
well-defined point nodes with the gap rising to slightly less that
one third of its maximum value between the nodes. When looked at
on the same scale as the $\gamma$ band, the $\beta$ band appears
to have line nodes (see Fig.~\ref{location2}). However, when
looked on a finer scale (see Fig.~ \ref{location3}), it is clear
that the nodes in the $\beta$ band are also point nodes, but that
gap on a line between nodes is roughly ten times smaller than it
is for the $\gamma$ band. Thus, there are lines of very small gap
for the $\beta$ band.

Our calculation of the sound attenuation will make use of the
electron-phonon matrix elements as determined in
Ref.~\protect\onlinecite{wak01}. As noted in the Introduction, the
extremely anisotropic nature of the observed ultrasonic
attenuation in Sr$_2$RuO$_4$ indicates \protect\cite{wak01} that
the modes denoted as L[100], L[110], and T[110] in
Ref.~\protect\onlinecite{lup01} are attenuated almost entirely by
interactions with $\gamma$-band electrons. (L and T mean
logitudinal and transverse, respectively, and the Miller indices
give the direction of propagation.)

The appropriate electron-phonon matrix element for the L[100] mode
is
\begin{equation}
    \label{long1}
            f^{L[100]}({\bk}) =  g^\gamma \cos(k_x a),
\end{equation}
whereas for longitudinal [110] waves the matrix element is
\begin{equation}
    \label{long2}
        f^{L[110]}({\bk}) = (g^\gamma/2) [\cos(k_x a)+\cos(k_y a)].
\end{equation}

For the transverse sound wave polarized in the basal plane and
propagating in the [110] direction, the matrix element describing
interactions with the $\gamma$-band electrons is
\begin{equation}
    \label{transv2}
        f^{T[110]}({\bk}) = (g^\gamma/2) [\cos(k_x a)-\cos(k_y a)]
\end{equation}
Notice that the matrix elements for the longitudinal modes are
nonzero at all of the point nodes in the $\gamma$ band indicated
in Fig.~\ref{nodes}.  Thus, all of the point nodes are ``active''
in the sense of Ref.~\protect\onlinecite{mor96}, as is generally
expected for longitudinal modes. Given that the low temperature
ultrasonic attenuation is expected to vary roughly as $T$ for line
nodes and as $T^2$ for point nodes (see Section III), the
experimentally observed temperature dependence\protect\cite{lup01}
of $T^{1.8}$ for the longitudinal nodes indicates that the
$\gamma$-band nodes are point nodes. Numerical studies confirm
this conclusion (see Section IV). Similar conclusions are reached
for the T[110] mode, but with the difference that its matrix
element is non-zero for the [100] nodes, but zero for the [110]
nodes.  This confirms the existence of [100] point nodes in the
$\gamma$ band.  According to the model discussed above, if [100]
nodes exist, then [110] nodes exist also.

The attenuation of the T[100] mode, which is extremely weak, must
be treated specially because, as described in
Ref.~\protect\onlinecite{wak01}, the dominant electron-phonon
interaction for $\gamma$-band electrons (described above in terms
of the interaction constant $g^\gamma$) is zero for T[100]
phonons.  Thus, the attenuation of this phonon mode must be due
either to a different, weaker, interaction with $\gamma$-band
electrons, e.g. described by the matrix
element\protect\cite{wak01}
\begin{equation}
    \label{transv1}
        f^{T[100]}({\bk}) = g'{}^\gamma \sin( k_x a ) \sin( k_y a ),
\end{equation}
or by an interaction with $\alpha$- and/or $\beta$-band electrons.
As described in Section IV, we have been unable to fit the
temperature dependence of the T[100] attenuation by assuming an
interaction of this mode with $\gamma$-band electrons.  On the
other hand, the attenuation of the T[100] mode can be accurately
fit by assuming that it interacts predominantly with electrons in
the $\alpha$ and $\beta$ bands.  For this to be the case it is
necessary to choose the model parameters so that the gap on a line
connecting $\alpha$- and/or $\beta$-band point nodes is relatively
small, so that at temperatures that are not too small the nodal
structure of the $\alpha$- and/or $\beta$-bands will appear
somewhat line-like.  This will provide consistency with a number
of
articles\protect\cite{mae02,nis99,yasu00,zhi01,wys03,miy99,deg04}
where line nodes are assumed to exist to account for the
thermodynamic properties of Sr$_2$RuO$_4$. The observed
temperature dependence for the T[100] attenuation is $T^{1.4}$,
which is in between the theoretical expectations for line and
point nodes, as might be expected for line-like point nodes.  An
excellent quantitative fit the the experimental data for the
T[100] mode, based on these ideas, is exhibited in Section IV.

\section{Calculation of the superconducting ultrasound attenuation}

This article uses a standard formalism, extended to account for
Fermi-surface and electron-phonon matrix-element anisotropy, to
calculate the ultrasonic attenuation.  This formalism is
summarized here, and described in further detail in the Appendix
of Ref.~\protect\onlinecite{wak01}.  A self-consistent calculation
of the gap function and of the energy-dependent impurity
scattering rate is beyond the scope of this article, which
considers the incorporation of anisotropic effects into the more
usual isotropic calculation as being the most essential for our
purposes.  Thus, in the superconducting state, we assume a
constant energy-independent quasiparticle lifetime equal to that
in the normal state.  In early studies of heavy-fermion
superconductivity, it was pointed out\protect\cite{var86} that, so
long as the temperature was not too low, such a quasiparticle
scattering rate could account for the ultrasound and heat
transport data in UPt$_3$. Similar results were found in
Refs.~\protect\onlinecite{wu02,gra02}. Therefore, the transport
and thermal properties of heavy-fermion superconductors can be
explained in terms of an effective electron scattering rate which
except for the lowest temperatures, is approximately temperature
independent and equal in magnitude to that of the normal state.

The expression that we use to calculate the ultrasonic attenuation
is an expression valid at low frequencies where that ultrasonic
attenuation is proportion to the square of the sound wave
frequency, as is the case experimentally\protect\cite{lup01}. This
expression, which is given in the appendix to
Ref.~\protect\onlinecite{wak01}, is
\begin{equation}
    \label{at}
        \frac{\alpha_j(\bq,T)}{\alpha_j(\bq,T_c)}=  \int_0^\infty
                d\epsilon \left(-\frac{\partial f}{\partial \epsilon}\right)
                        \frac{A^i_j(\bq,\epsilon)}{\epsilon},
\end{equation}
where
\begin{equation}
    \label{fat}
            A^i_j(\bq,\epsilon)=  \frac{\left\langle
                    \tilde{f}_{j,\gamma}^2(\bk,\bq)\re\sqrt{\epsilon^2-|\Delta^{\gamma}_\bk|^2}
                        \right\rangle_{FS}}{\left\langle
                                \tilde{f}_{j,\gamma}^2(\bk,\bq)\right\rangle_{FS}}.
\end{equation}
and where $j$ indicates the phonon mode, and $\left< \right>_{FS}$
indicates a Fermi-surface average. It should be noted that, in
order to preserve charge neutrality in the distorted lattice in
the presence of a longitudinal sound wave \protect\cite{akh57},
the electron-phonon matrix elements of the previous section must
be replaced by effective matrix elements defined by
\begin{equation}
    \label{ep_eff}
            {\tilde f}^j(\bk) = f^j(\bk) - \left<f^j(\bk)\right>_{FS}.
\end{equation}
in the formula for the ultrasonic attenuation.

Since we have not carried out a self-consistent evaluation of the
gap, we do not have a model for the temperature dependence of the
functions $\Delta^i (T)$ appearing in Eq.~\ref{sup_spectrum}. In
the absence of such a model, we simply assume a temperature
dependence of the form $\Delta^i(T)$ = $\Delta_0^i~\sqrt{1 -
(T/T_c)^3}$, which is sometimes used in the literature (e.g. see
Ref.~\protect\onlinecite{wu02}). However, the choice of
$\Delta(T)$ does not seem to affect our results, in particular we
did not see any difference in the fits between $(T/T_c)^3$ and
$(T/T_c)^2$.

It is useful for an initial interpretation of the ultrasonic
attenuation data to have a simple expression for the expected
temperature dependence of the attenuation at low temperatures for
both line nodes and point nodes.  For nodes that are ``active''
for a given phonon mode, the electron-phonon matrix element for
that mode is by definition non-zero at the node and can be
approximated by a constant, independent of wave vector, in the
low-temperature limit. In this case, Eqs.~\ref{at} and \ref{fat}
show that the ultrasonic attenuation in the low-temperature limit
varies as
\begin{equation}
        \label{usLine}
            \alpha = C \; T
\end{equation}
for line nodes, and
\begin{equation}
    \label{usPoint}
            \alpha = C^\prime  \; T^2
\end{equation}
for point nodes, where $C$ and $C^\prime$ are constants
independent of temperature. For nodes that are inactive for a
given phonon mode, these temperature dependences are one power of
$T$ higher. \protect\cite{mor96}

\section{Quantitative results}

\begin{figure}
\includegraphics[width = 3.0 in, height= 4.8 in]{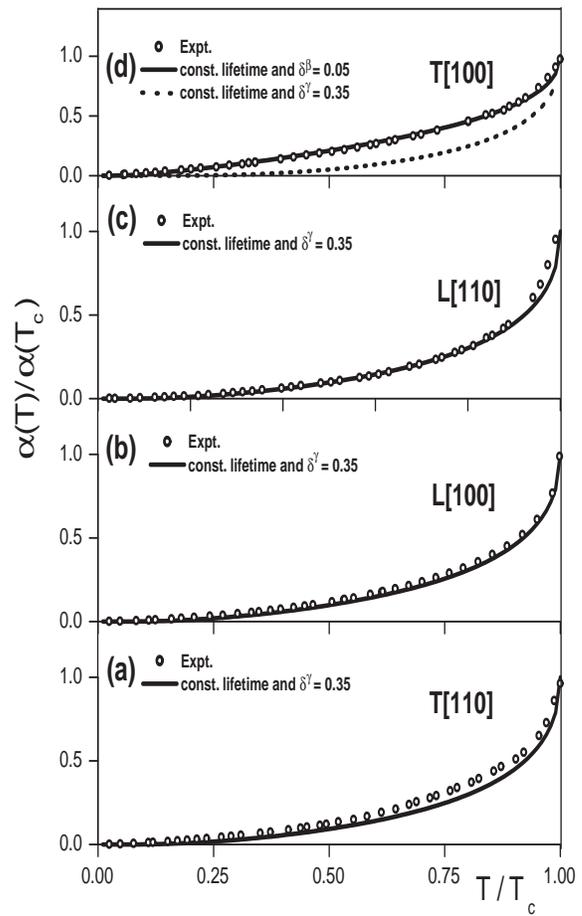}
\caption{\label{fits} Numerical fits of the in-plane ultrasound
attenuation modes T[110], L[100], L[110], and T[100] normalized at
$T_c$. Experimental data is taken from Ref.~\protect\cite{lup01}.
The free parameters are $\delta$ and $\Delta_0$. The lifetime used
to evaluate the expression for the attenuation is taken to be
constant and equal to the one in the normal state.}
\end{figure}

Fig.~\ref{fits} shows the theoretical results for the temperature
dependence of normalized ultrasonic attenuation calculated by
evaluating Eq.~\ref{at}, as well as the experimental results from
Ref.~\protect\onlinecite{lup01}, for the T[100],T[110], L[100],
and L[110] modes. In obtaining the theoretical results, the
parameters $\Delta^i_0$ and $\delta^i$ for $i = \gamma, \beta$
were determined in such a way as to give the best fit to the data.

First, the fitting of the data for the three relatively strongly
attenuated phonon modes (L[100], L[110] and T[110]) [shown in
panels (a), (b) and (c)] was attempted. The parameters
$\Delta^\gamma_0 = 0.7$ meV and $\delta^\gamma = 0.35$ gave
excellent fits for the temperature dependence of all three modes.
(The value of $\Delta^\gamma$ obtained here is in agreement with
that maximum gap found in a recent scanning tunnelling microscopy
experiment\protect\cite{upw01}, showing at least that it has a
reasonable order of magnitude.) The strong anisotropy in the
attenuation (the mode viscosity of the L[110] mode is a factor of
30 smaller that that for the L[100] and T[110] modes) is not
apparent in this figure since the attenuation relative to that at
$T=T_c$ is plotted for each mode. It should be emphasized,
however, that this anisotropy has been shown to be accurately
accounted for by the assumed electron-phonon interaction, which
(for these three modes) is characterized by a single constant, see
Eqs. (\ref{long1}), (\ref{long2}), (\ref{transv2}). This, as well
as the excellent agreement between the theoretical and
experimental temperature dependence for all three modes exhibited
in Fig.~\ref{fits}, provides strong support for the essential
features of the proposed model.

As a final step, we attempted to fit the temperature-dependent
attenuation of the most weakly attenuated mode, the T[100] mode.
As noted in Section II, the largest component of the
electron-phonon interaction, which is responsible for the
attenuation of the three modes discussed in the preceding
paragraph, is exactly zero for the T[100] mode. Thus it is
necessary to determine which new secondary electron-phonon
interaction is appropriate for describing the attenuation of the
T[100] mode, and in particular to which band the electrons
contributing the most to this attenuation belong.  (The theory of
the different possible electron-phonon interactions is described
in detail in Ref.~\protect\onlinecite{wak01}). As a first attempt,
the temperature-dependent attenuation was calculated using the
electron-phonon matrix element of Eq.~\ref{transv1} and assuming
that the dominant interaction is with electrons in the $\gamma$
band. The result, shown in Fig.~\ref{fits}(d), gives a very poor
fit to the experimental data.  This should perhaps have been
expected since the attenuation of the T[100] mode varies roughly
as $T^{1.4}$ at low temperatures, in comparison with $T^{1.8}$ for
the other three modes. Now, according to the theory, the limiting
low-temperature temperature dependences should be determined by
the nodal structure, so it is reasonable to assume that the
attenuation of the T[100] modes is by its interaction with
electrons in another band and with a different nodal structure,
say the $\alpha$ or $\beta$ band.  Thus, we use the same model as
used for the other three modes, except that we allow the parameter
$\delta^i$ (which here we call $\delta^\beta$) in
Eqs.~\ref{real_gap} and \ref{imag_gap} to vary so as to achieve
the best fit with the data.  The best value of this parameteris
$\delta^\beta = 0.05$, and also $\Delta^\beta_0 = 0.7$ meV. The
nodal structure of the $\beta$ band, calculated using this value
of $\delta^\beta$, is shown above in Figs.~\ref{location2} and
\ref{location3}.  Note that, although the nodes are point nodes,
they have some of the properties of line nodes as there is a very
low gap along the line joining the nodes. Hence, very roughly,
this result suggests a temperature dependence for the attenuation
somewhere between $T$ (appropriate for line nodes) and $T^2$
(appropriate for point nodes) as is indeed the case.

\section{conclusions}

A detailed quantitative interpretation of the
temperature-dependent ultrasonic attenuation of Sr$_2$RuO$_4$
leads to the determination of essentially different nodal
structures on two different bands of this multi-band
superconductor. This remarkable determination is possible because
different phonon modes interact most strongly with electrons in
different bands in this material.  Furthermore, the exceptionally
strong anisotropy in the attenuation of certain modes, which is
unique to this material, allows one to associate the attenuation
of the most strongly attenuated modes with their interaction with
electrons in the $\gamma$ band. Thus, the nodal structure of the
$\gamma$ band is found to be characterized by at least eight
well-defined point nodes, symmetrically distributed in \{100\}
planes, and also by another eight point nodes, symmetrically
distributed in \{110\} planes. The attenuation of the most weakly
attenuated mode gives information about the nodal structure of the
$\alpha$ and $\beta$ bands. The $\alpha / \beta$ band nodal
structure is characterized by the existence of eight point nodes,
symmetrically distributed in \{110\} planes, in either or both of
the $\alpha$ and $\beta$ bands, and also by another eight point
nodes symmetrically distributed in \{100\} planes, in either or
both of the $\alpha$ and $\beta$ bands. The nodal structure here
is, however, quantitatively different from that of the $\gamma$
band, in that the gap on a line on the Fermi surface joining the
nodes is an order of magnitude smaller than that on the $\gamma$
band. This gives the $\alpha$- and/or $\beta$-band nodal
structures some of the characteristics of line nodes.

\vspace{0.08cm}
\section{acknowledgments}

We thank L. Taillefer, M. Tanatar, I. Mazin, and W. Kim  for
stimulating discussions, and C. Lupien for providing us with the
experimental data. We also thank the Canadian Institute for
Advanced Research and the Natural Sciences and Engineering
Research Council of Canada for their support. MBW thanks Efim Kats
and the Theory Group of the Institute Laue Langevin (where part of
this work was carried out) for their hospitality.

\end{document}